\documentclass[aps,pra,preprint,groupedaddress,showkeys,showpacs]{revtex4}

\usepackage{graphicx}
\usepackage{dcolumn}
\usepackage{bm}
\usepackage{amsmath}
\usepackage{amssymb}
\usepackage{latexsym}
\usepackage{epsfig}
\usepackage{amsbsy}
\usepackage{array}
\usepackage{amssymb}
\usepackage{setspace}
\usepackage{bm}
\usepackage{epstopdf}
\usepackage{slashed}
\usepackage{multirow}
\usepackage[flushleft]{caption}
\def\sint{\ifmmode{- \!\!\!\!\!\! \int}
    \else{\hbox{$- \!\!\!\! \int \ $}}\fi}

\begin{document}


\title{Searches for the Anomalous FCNC Top-Higgs Couplings with Polarized  Electron Beam at the LHeC}

\author{XiaoJuan Wang,
Hao Sun\footnote{Corresponding author: haosun@mail.ustc.edu.cn \hspace{0.2cm} haosun@dlut.edu.cn} and Xuan Luo}
\affiliation{\footnotesize Institute of Theoretical Physics, School of Physics $\&$ Optoelectronic Technology,
 Dalian University of Technology, No.2 Linggong Road, Dalian, Liaoning, 116024, P.R.China}
 

\begin{abstract}
In this paper, we study the single top and Higgs associated production
$\rm e^- p\rightarrow \nu_e \bar{t} \rightarrow \nu_e h \bar{q}(h\rightarrow b\bar{b})$ 
in the top-Higgs FCNC couplings at the LHeC with the electron beam energy of $E_{e}$ = 60 GeV 
and $E_{e}$ = 120 GeV, combination of a 7 TeV and 50 TeV proton beam. With the
possibility of e-beam polarization ($p_{e}$ = 0, $\pm0.6$), we distinct the Cut-based method 
and the Multivariate Analysis (MVA) based method, and compare with the current 
experimental and theoretical limits. It is shown that the branching ratio 
 Br $\rm(t\to uh)$ can be probed to 0.113 (0.093) $\%$, 0.071 (0.057) $\%$, 0.030 (0.022) $\%$ and
 0.024 (0.019) $\%$ with the Cut-based (MVA-based) analysis 
at ($E_{p}$, $E_{e}$) = (7 TeV, 60 GeV), ($E_{p}$, $E_{e}$) = (7 TeV, 120 GeV), ($E_{p}$, $E_{e}$) = (50 TeV, 60 GeV) and ($E_{p}$, $E_{e}$) = (50 TeV, 120 GeV) beam
energy and 1$\sigma$ level. With the possibility of
e-beam polarization, the expected limits can be probed down to 
0.090 (0.073) $\%$, 0.056 (0.045) $\%$, 0.024 (0.018) $\%$ and 0.019 (0.015) $\%$, respectively. 
\end{abstract}

\pacs{14.65.Ha, 12.60.-i}

\keywords{Top quark, Higgs boson, Anomalous Couplings, LHeC, Polarization}

\maketitle

\section{Introduction}  
The Large Hadron Electron Collider (LHeC) is the second electron-hadron collider following HERA \cite{LHeC_design_Note}. 
With remarkable higher energy and luminosity, the LHeC is a major step towards understanding the Higgs physics and QCD. 
For the LHeC colliding energy, the 7 TeV proton beam at the LHC as well as the 50 TeV proton beam at the future FCC-he \cite{FCC-he_Study}
and a new 60 GeV electron beam \cite{LHeC_design_Note} are envisaged. To probe new physics, the anomalous flavor changing neutral current (FCNC) Yukawa interactions, between the top-Higgs and either an up or charm quark,
would provide a clear signal. The SM Lagrangian can be extended
by the following terms,\begin{eqnarray}\label{lagrangian}
\rm {\cal L} = \kappa_{tuh} \bar{t}uh + \kappa_{tch} \bar{t}ch + h.c.,
\end{eqnarray}
where the real parameters $\rm \kappa_{tuh}$ and $\rm \kappa_{tch}$ denote the FCNC couplings of the  Higgs to up-type quarks.
The total decay width of the top-quark $\Gamma_t$ is 
\begin{eqnarray}
\rm \Gamma_t &=&\rm  \Gamma^{SM}_{t\rightarrow W^-b}+\Gamma_{t\rightarrow ch}+\Gamma_{t\rightarrow uh}. 
\end{eqnarray}
where the decay width $\rm \Gamma_{t\to W^-b}^{SM}$ and $\rm \Gamma_{t\rightarrow u(c)h}$ can be found in \cite{twb_NLO} and
\cite{decay_tqH}, respectively. Thus, the branching ratio for  $\rm t \rightarrow u(c)h$ can be approximately given by 
\begin{eqnarray}
\rm Br(t\rightarrow u(c)h) =
\frac{\kappa_{tu(c)h}^2}{\sqrt{2} G_F m_t^2 } \frac{(1-\tau_h^2)^2}{(1-\tau_W^2)^2(1+2 \tau_W^2)}\approx0.512{\kappa_{tu(c)h}^2}
\end{eqnarray}
where $\rm G_F$ is the Fermi constant and $\rm \tau_W=\frac{m_W}{m_t}$. The W boson and top quark masses are chosen to be 
$\rm m_W = 79.82\ GeV$ and $\rm m_t = 173.2\ GeV$, respectively.

Up to now, the investigation of $\rm t\to qh$ 
anomalous couplings have been experimented by many groups, which gives the stronge limits on the top-Higgs FCNC couplings. For 
instance, according to the ATLAS and CMS collaborations, the upper limits of  Br $\rm(t\to qh)<0.79$ $\%$ \cite{FCNC_limit_tHq_ATLAS,FCNC_limit_tHq_ATLAS1}
and  Br $\rm(t\to qh)<0.45$ $\%$ \cite{FCNC_limit_tHq_CMS} have been set at 95 $\%$ confidence level (C.L.). Except for the direct collider 
measurements, the low energy observable, by bounding the tqH vertex from the observed $\rm D^0-\bar{D^0}$ mixing \cite{FCNC_limit_tHq_DDbar},
the upper limit of  Br $\rm(t\to qh)<5\times 10^{-3}$ may be produced. Furthermore, through $\rm Z\to c\bar{c}$ decay and 
electroweak observables, the upper limit of  Br $\rm(t\to qh)<0.21$ $\%$ \cite{tqh_z2bb} can be obtained.

On the other hand, based on the experimental data, many phenomenological studies are performed from different channels.
For instance, \cite{tqh_ppWhj} found that the branching ratios Br $\rm(t\to qh)$ can be probed to 0.24 $\%$ at 3$\sigma$ level 
at 14 TeV LHC with an integrated luminosity of 3000 $\rm fb^{-1}$ through the process $\rm Wt\to Whq \to \ell\nu b\gamma\gamma q$.
\cite{EFT_tqH_2} explored the top-Higgs FCNC couplings through $\rm t\bar{t}\to Wbqh\to \ell\nu b\gamma\gamma q$ and 
found the branching ratios Br $\rm(t\to uh)$ can be probed to 0.23 $\%$ at 3$\sigma$ sensitivity at 14 TeV LHC with L = 3000 $\rm fb^{-1}$.
And \cite{ILC_tqh} obtained the Br $\rm(t\to qh)$ to be 0.112 $\%$ based on the process of $\rm t\bar{t}\to tqh\to \ell\nu bb\bar{b} q$.
The process of $\rm th\to \ell\nu b\tau^+\tau^-$ has been studied in \cite{EFT_tqH_3} and they estimated the upper limits of
Br $\rm(t\to uh)<0.15$ $\%$ at 100 $\rm fb^{-1}$ of 13 TeV data for multilepton searches. The results from different experiments and theoretical 
channels are summarized in Table \ref{SBaftercuts_1}.
\begin{table}[htbp]

 \caption{\label{SBaftercuts_1}
The results from different experimental and phenomenological channels.}
\begin{center}
\begin{tabular}{c| c |c }
\hline\hline
Channels      &Data Set &Limits  \\
\hline
$\rm t\bar{t}\to Wbqh \to \ell\nu b\gamma\gamma q$  &  ATLAS, 4.7 (20.3) $\rm fb^{-1}$ @ 7 (8) TeV &  $\rm Br$ $(t\to qh)< 0.79$ $\%$ \cite{FCNC_limit_tHq_ATLAS, FCNC_limit_tHq_ATLAS1}   \\
\hline
$\rm t\bar{t}\to Wbqh\to \ell\nu b\gamma\gamma q$ & CMS, 19.5 $\rm fb^{-1}$ @ 8 TeV     &$\rm Br$ $(t\to uh)< 0.45$ $\%$ \cite{FCNC_limit_tHq_CMS}        \\
\hline
$\rm D^0-\bar{D^0}$ mixing data  &    -      &   $\rm Br$ $(t\to qh)<0.5$ $\%$ \cite{FCNC_limit_tHq_DDbar}           \\
\hline
$\rm Z\to c\bar{c}$ and EW observables&     -       &    $\rm Br$ $(t\to qh)<0.21$ $\%$\cite{tqh_z2bb}       \\
\hline
$\rm Wt\to Whq \to \ell\nu b\gamma\gamma q$ & LHC, 3000 $\rm fb^{-1}$ @ 14 TeV, $3\sigma$  &   $\rm Br$ $(t\to qh)< 0.24$ $\%$ \cite{tqh_ppWhj}         \\
\hline
$\rm t\bar{t}\to Wbqh\to \ell\nu b\gamma\gamma q$  &  LHC, 3000 $\rm fb^{-1}$ @ 14 TeV       &   $\rm Br$ $(t\to uh)< 0.23$ $\%$ \cite{EFT_tqH_2}       \\
\hline
$\rm t\bar{t}\to tqh\to \ell\nu bb\bar{b} q$ &   ILC, 3000 fb$^{-1}$ @ 500 GeV      &   $\rm Br$ $(t\to qh)< 0.112$ $\%$ \cite{ILC_tqh}        \\
\hline
$\rm th\to \ell\nu b\tau^+\tau^-$ &LHC, 100 fb$^{-1}$ @ 13 TeV     &$\rm Br$ $(t\to uh)< 0.15$ $\%$ \cite{EFT_tqH_3}  \\
\hline
$\rm th\to \ell\nu b\ell^+\ell^-X$ &LHC, 100 fb$^{-1}$ @ 13 TeV &$\rm Br$ $(t\to uh)< 0.22$ $\%$ \cite{EFT_tqH_3} \\
\hline
$\rm th\to jjb b\bar{b}$ &LHC, 100 fb$^{-1}$ @ 13 TeV &$\rm Br$ $(t\to uh)< 0.36$ $\%$ \cite{EFT_tqH_3}\\
\hline\hline
 \end{tabular}
 \end{center}

\end{table}

\begin{figure}[hbtp]

\includegraphics[scale=0.15]{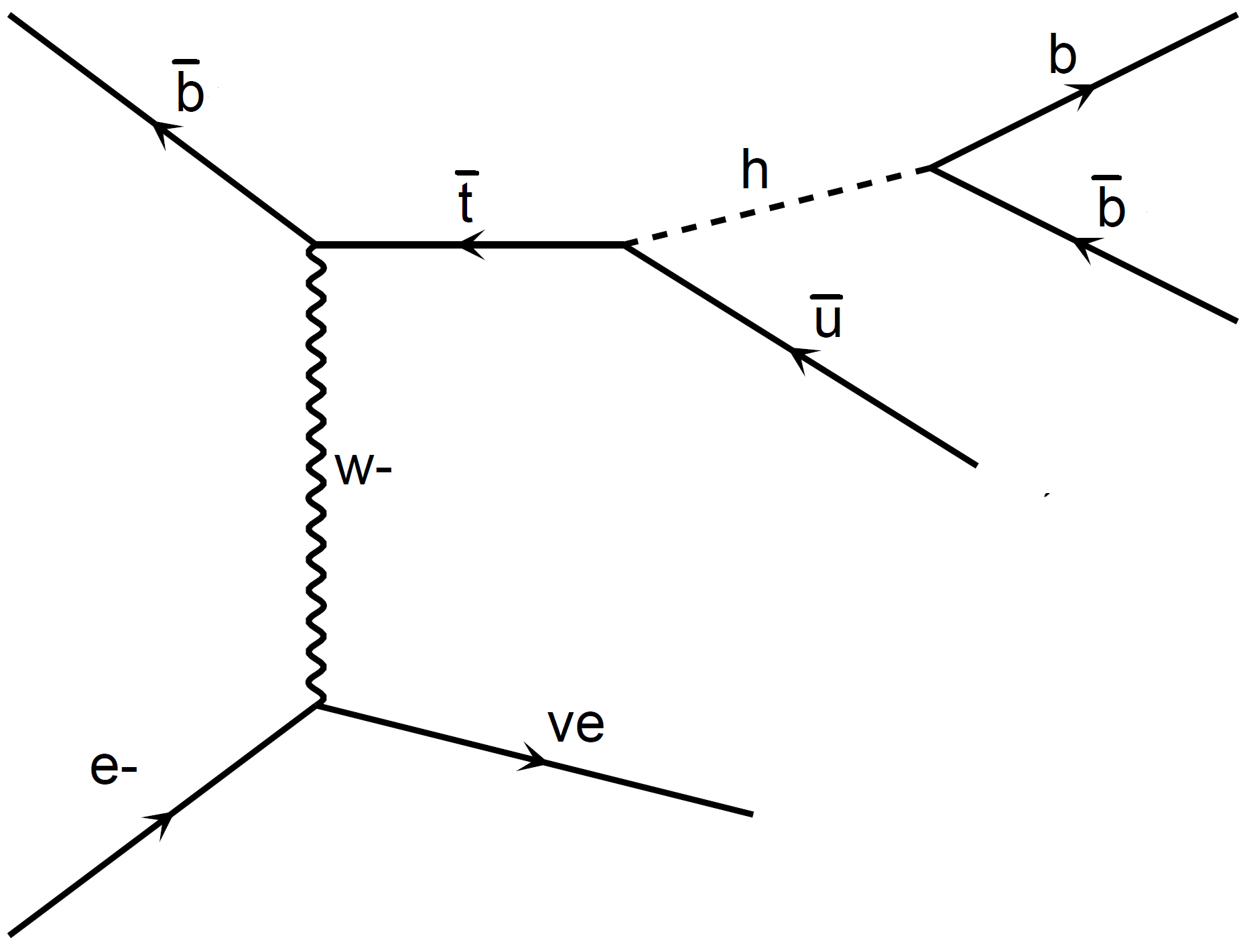}
\caption{\label{fig1_sig_Feynman}
Feynman diagram for the partonic process $\rm e^{-} \bar{b} \to \nu_e \bar{t} \rightarrow \nu_e h \bar{q} \to \nu_e b\bar{b} \bar{q}$ 
at the LHeC through Flavor Changing top-Higgs interactions.}
\end{figure}

In this study, we examined the $\rm e^- p\rightarrow \nu_e \bar{t} \rightarrow \nu_e h \bar{q}$ at the LHeC where the Higgs boson decays
to $b\bar{b}$, at a 7 (50) TeV with a 60 (120) GeV electron beam and 1000 $\rm fb^{-1}$ integrated luminosity. The possibility of e-beam polarization
is also considered. The Feynman diagram is plotted in Fig. \ref{fig1_sig_Feynman}. The main backgrounds which yield the same or similar
final states to the signal are listed as below:
\begin{eqnarray}\nonumber
&&\rm e^-p \rightarrow \nu_e (\bar{t} \rightarrow ( W^- \rightarrow jj) \bar{b} ) \\\nonumber
&&\rm e^-p \rightarrow e^- jjj  \\\nonumber
&&\rm e^-p \rightarrow \nu_e jjj  \\\nonumber
&&\rm e^-p \rightarrow \nu_e (h\rightarrow b \bar{b} ) j  \\
&&\rm e^-p \rightarrow \nu_e (z\rightarrow b \bar{b} ) j,
\end{eqnarray}
where j = g, u, $\rm \bar{u}$, d, $\rm \bar{d}$, c,
$\rm  \bar{c}$, s, $\rm \bar{s}$, b and $\rm \bar{b}$ if possible.
Notice $\rm e^-p \rightarrow e^- jjj$ is the neutral current  multi-jet QCD background, and all  the others are 
belong to charged current (CC) productions. For the single top background $\rm e^-p \rightarrow \nu_e (\bar{t} \rightarrow ( W^- \rightarrow jj) \bar{b} )$,
the produced top quark will decay to a W boson and a b-jet. The W boson continues to decay to non-b-jet final states, which might mis-tagged
as a b-jet. With the same final states, $\rm e^-p \rightarrow \nu_e (h\rightarrow b \bar{b} ) j$ and $\rm e^-p \rightarrow \nu_e (z\rightarrow b \bar{b} ) j$
are the irreducible backgrounds corresponding to associated Higgs jet and Z jet which contain three QED couplings. $\rm e^-p \rightarrow \nu_e jjj$ is 
the CC multi-jet QCD background. Similar as  the single top background, a mis-identification of one or more of the final state light jets to b-jet, makes this process  a reducible background.

\section{Tools and Method}
During the simulation, we first extract the Feynman Rules by using the FeynRules package \cite{FeynRules2.0} and generate the event 
with MadGraph@NLO \cite{MadGraph5}. PYTHIA6.4 \cite{Pythia6.4} was set to solve the initial and final state parton shower, hadronization,
heavy hadron decays, etc. We use CTEQ6L \cite{CTEQ6L} as the parton distribution function and set the renormalization and factorization
scale to be $\mu_{r}=\mu_{f}$. We take the input heavy particle masses as $\rm m_h=125.7\ GeV$, $\rm m_t=173.2\ GeV$,
$\rm m_Z=91.1876\ GeV$ and $\rm m_W=79.82\ GeV$, respectively. We employ the following basic pre-selections cuts to select the 
events:
\begin{eqnarray}\label{generator-level-cuts}\nonumber
&&\rm  \slashed E^{missing}_T \geq 15\ GeV, \\\nonumber
&&\rm   p_T^{k_0} \geq 15\ GeV, \ \ k_0=j, b, \ell,  \\\nonumber
&&\rm  |\eta^j| < 5, |\eta^b|<5, |\eta^\ell| \leq 3, \\
&&\rm  \Delta R(k_1k_2) >0.4, \ \ k_1k_2=jj, j\ell, jb, bb, b\ell.
\end{eqnarray}
where $\rm \Delta R = \sqrt{\Delta \Phi^2 + \Delta \eta^2}$ is the separation with $\Delta \eta$ and $\Delta \Phi$
in the rapidity-azimuth plane, $\rm p_T^{jet, b, \ell}$ and $\rm  |\eta^{jet,b, \ell}|$
are the transverse momentum and the pseudo-rapidity of jets, b-jets and leptons
while $\rm  \slashed E_T^{missing}$ is the missing transverse momentum. Then we adopt a  Cut-based method and a Multivariate Analysis (MVA) based method 
for signal and background analysis, respectively.

\subsection{Cut-based method}
In order to distinguish between signal-related events and background-related
events as much as possible, we set a series of cuts. We list all the Cut-based selections here:
\begin{itemize}
\item cut1: the basic pre-selection cuts.
\item cut2: the selection $\rm  e^- p \rightarrow \slashed{E}^{missing}_T + 0\ \ell + \geq 3\ jets,\ (with\ at\ least\ 2\ tagged\ b-jets).$
\item cut3: Missing transverse energy $\rm {\slashed E}^{missing}_T > 20\ GeV$.
\item cut4: the reconstructed top quark mass window $\rm m_t \in [148\ GeV, 178\ GeV]$.
\item cut5: the reconstructed W boson mass window $\rm m_W <50\ GeV\ or\ m_W>90\ GeV$.
\item cut6: the reconstructed Z boson mass window $\rm m_Z<55\ GeV\ or\ m_Z>95\ GeV$.
\item cut7: the reconstructed higgs mass window $\rm m_h \in [100\ GeV, 130\ GeV]$.
\end{itemize}
\subsection{MVA-based method}
We implemented the MVA method using the Root Toolkit for Multivariate Analysis (TMVA) \cite{TMVA}.
After cut1, cut2 and cut3, we especially select several input variables 
to discriminate the signal and background events, thus resulting better signal significance. Specifically, we define 
a set of totally 44 kinematic variables and choose the most effective ones for Boosted Decision Trees (BDT) training,
which are: the b-jet number ($\rm N_{bjet}$), the separation in the $\rm \Phi-\eta$ plane
between jets ($\rm \Delta R^{B_1B_2}$,  $\rm \Delta R^{B_1J_1}$), the difference in azimuthal angle between jets ($\rm \Delta \Phi^{B_1B_2}$,
$\rm \Delta \Phi^{B_1J_1}$), the transverse momentum of the jet
($\rm p_{T}^{J_1}$), the difference in $|\eta|$ between Higgs jet system ($\rm \Delta \eta^{hJ_1}$).
It is worth noting that e-beam polarization is 
considered in both Cut-based method and MVA-based method.
\section{Results}
\begin{figure}[hbtp]
\centering
\includegraphics[scale=0.27]{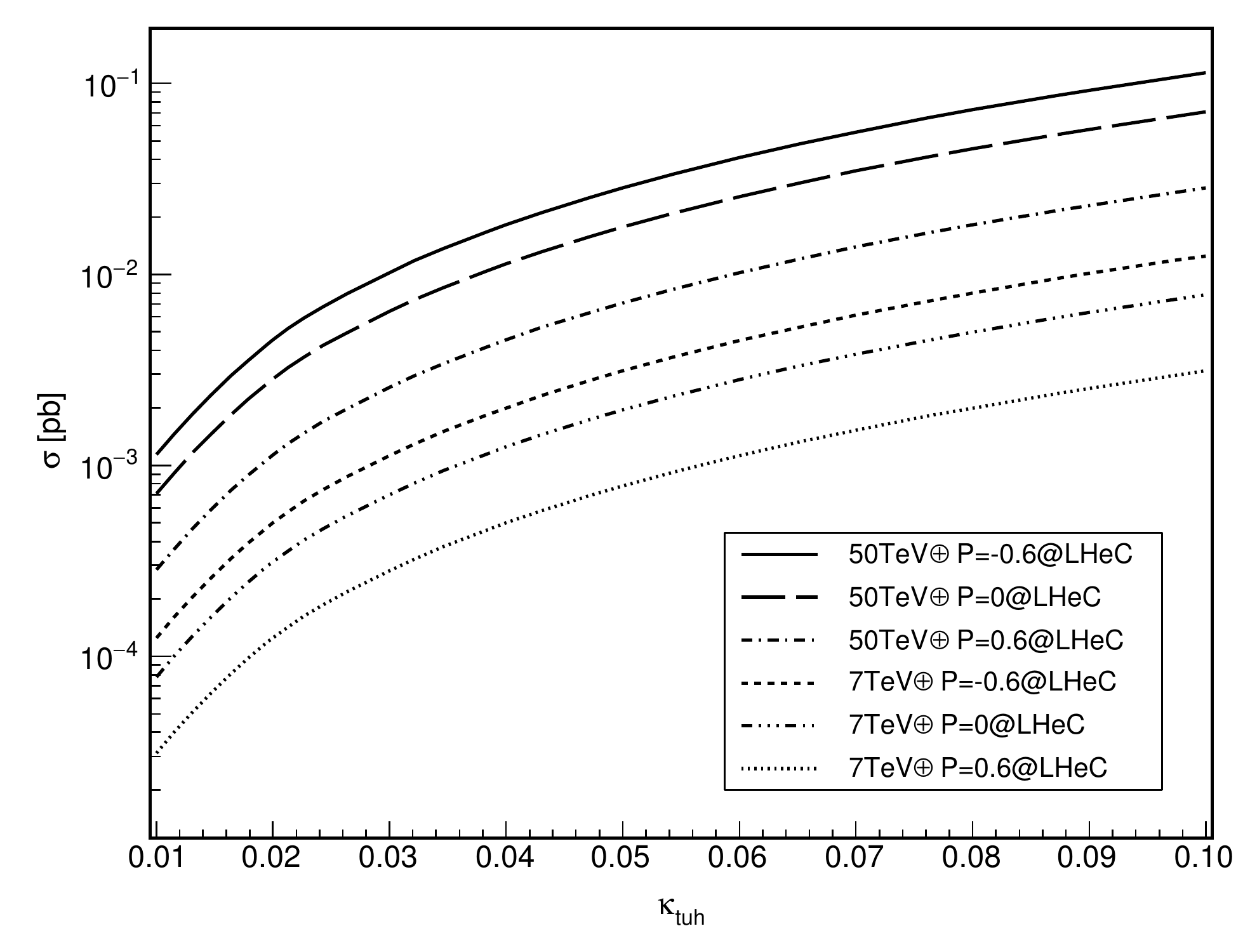}
\includegraphics[scale=0.27]{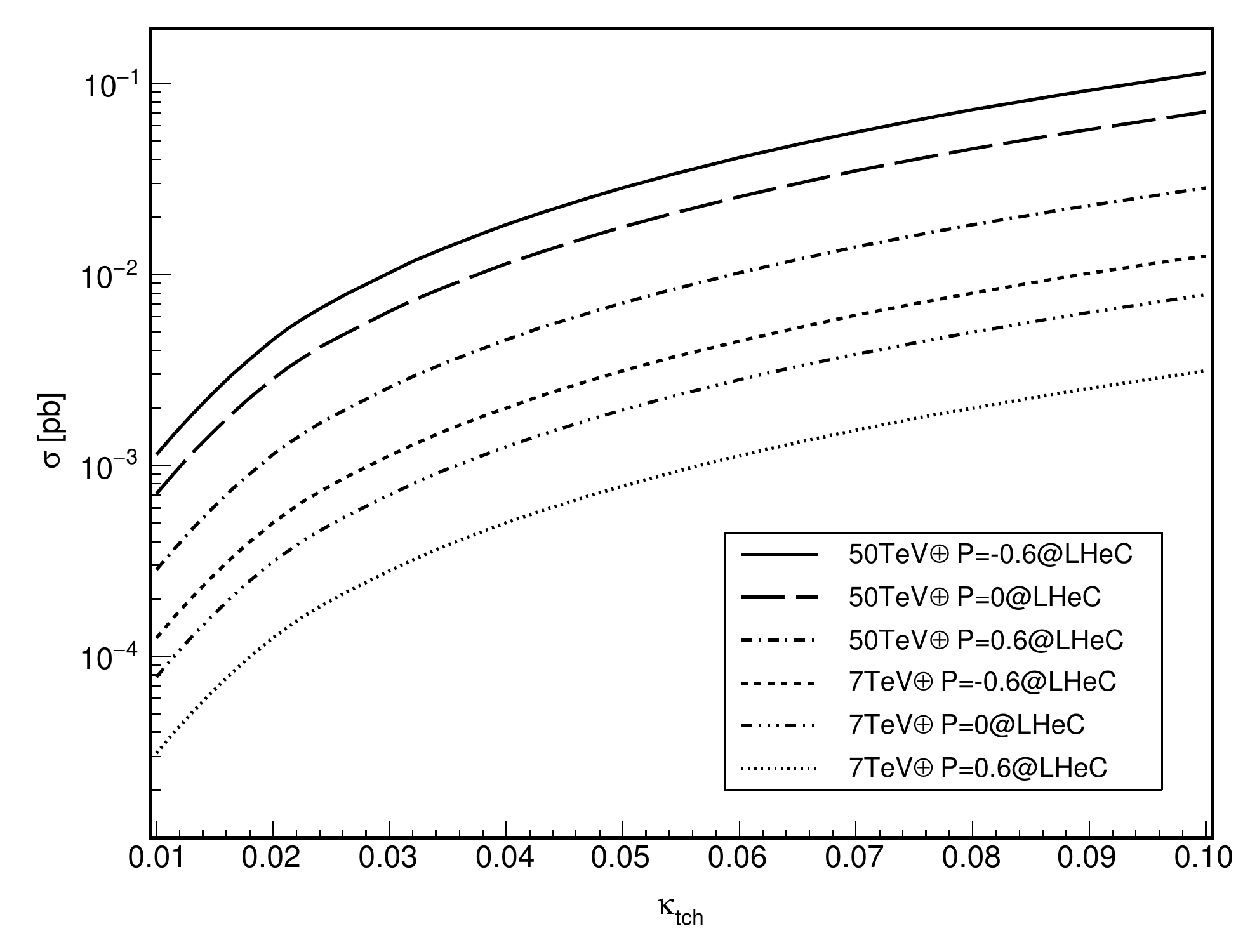}
\includegraphics[scale=0.27]{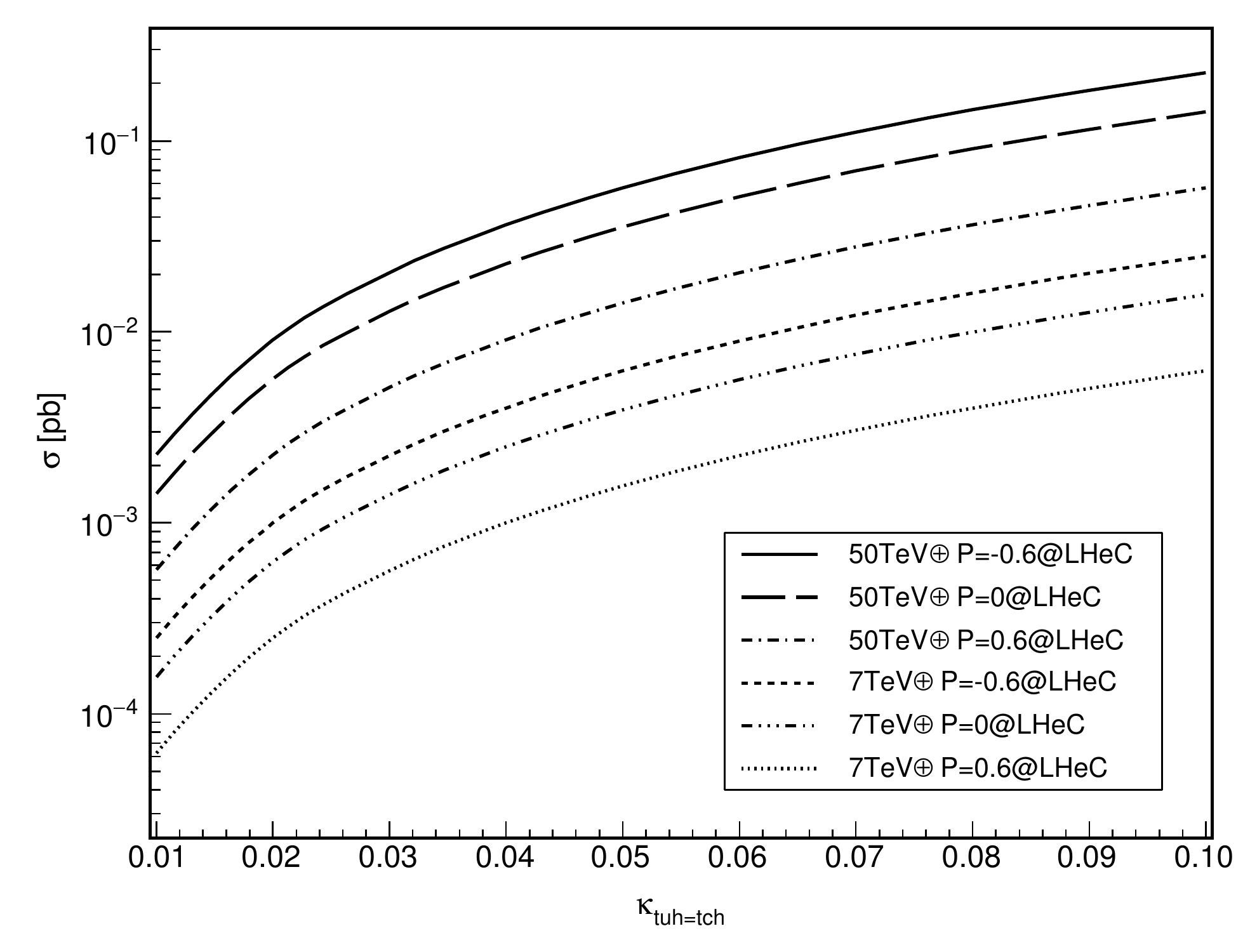}
\caption{\label{fig3_kqh_60}
The cross sections $\rm \sigma_{tqh}$ on the top-Higgs FCNC
couplings $\rm \kappa_{tqh}$ at the 7 (50) TeV and 60 GeV LHeC with e-beam polarization $p_{e}=0,\pm0.6$.}
\end{figure}

\begin{figure}[hbtp]
\centering
\includegraphics[scale=0.27]{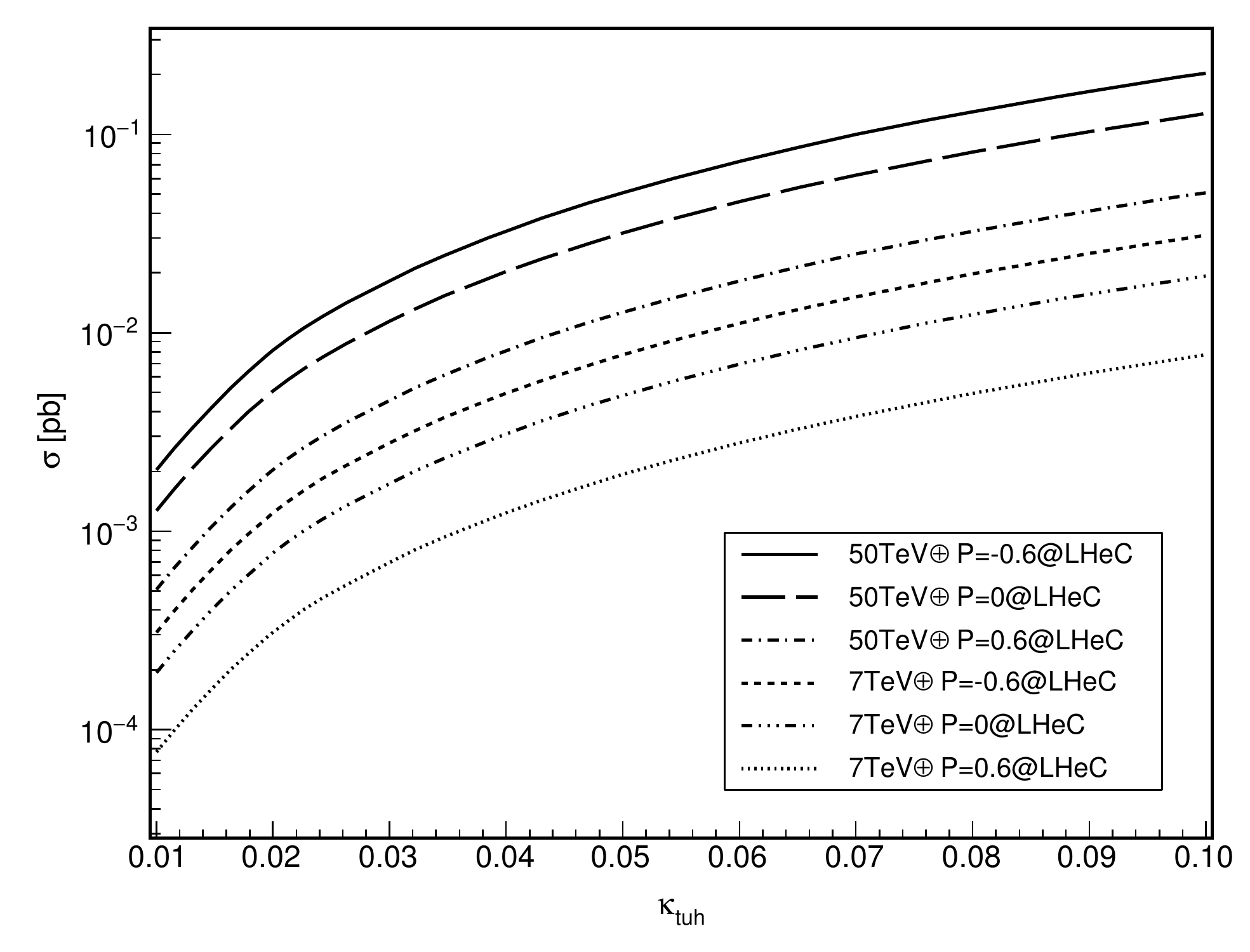}
\includegraphics[scale=0.27]{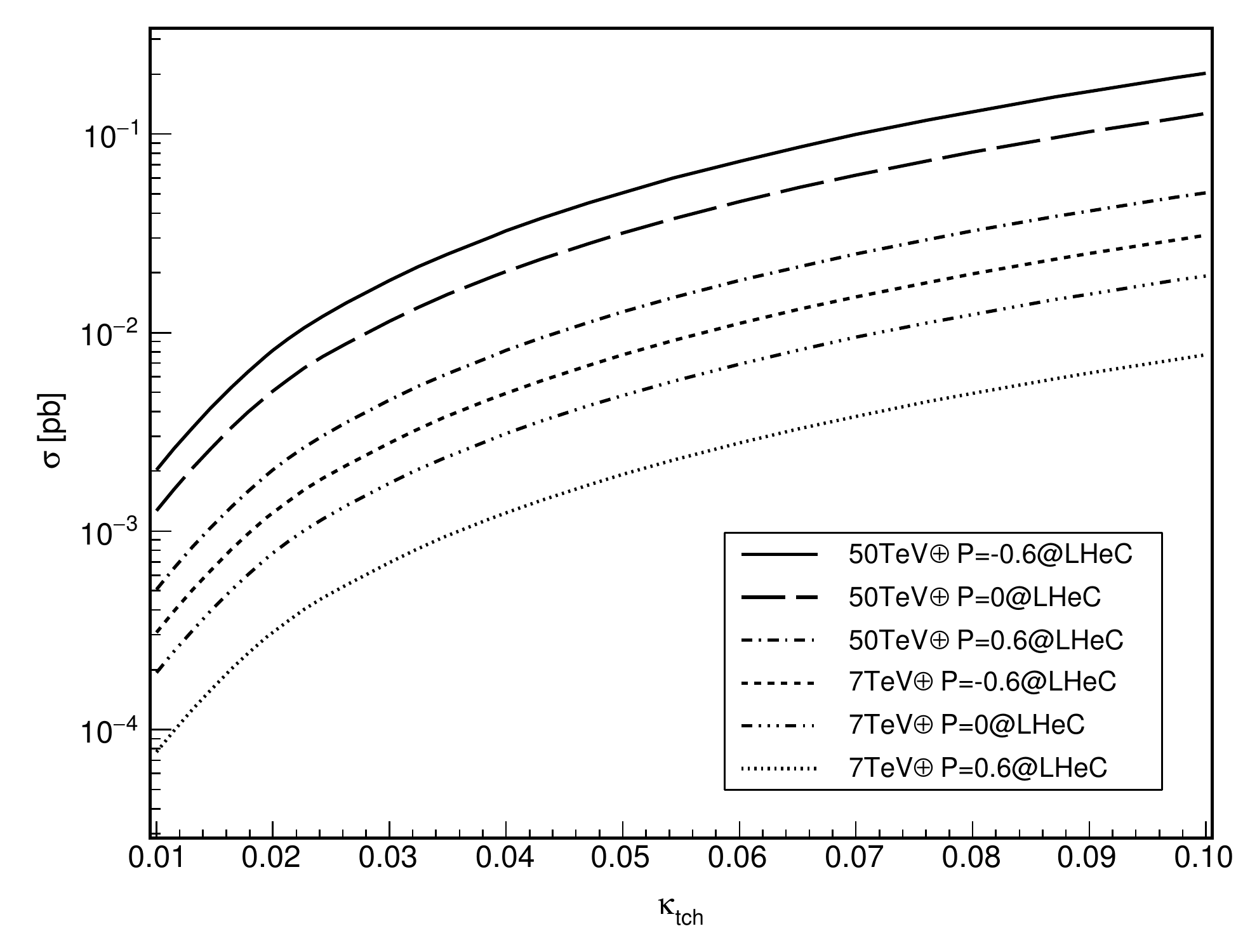}
\includegraphics[scale=0.27]{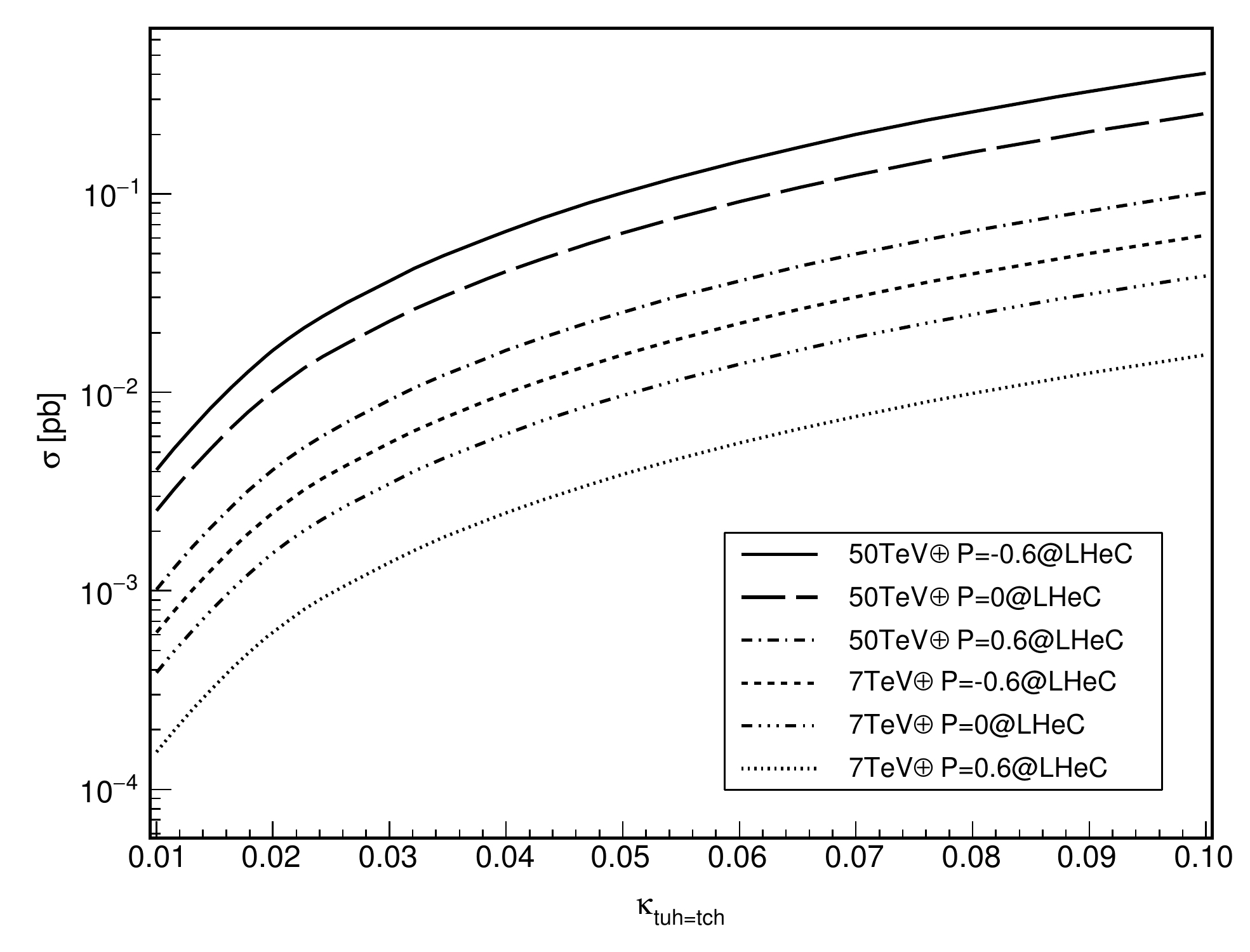}
\caption{\label{fig4_kqh_120}
The same as Fig. \ref{fig3_kqh_60} but for $E_{e}$ = 120 GeV. }
\end{figure}

In Fig. \ref{fig3_kqh_60} (60) GeV and Fig. \ref{fig4_kqh_120} (120) GeV, we show the dependence of the cross section 
$\rm \sigma$ on the top-Higgs FCNC couplings  $\rm \kappa_{tqh}$ at $E_{e}$ = 60 (120) GeV with $p_{e}$ = $\pm$0.6 electron
beam polarization combination of a 7 (50) GeV proton beam for three different cases. (I) $\kappa_{tqh}= \kappa_{tuh}, \kappa_{tch} =0, $
(II)  $\kappa_{tqh}= \kappa_{tch}, \kappa_{tuh} =0 $ and (III) $\kappa_{tqh}= \kappa_{tuh} = \kappa_{tch} $. Obviously,
the cross section of $\kappa_{tqh}=0.1$ can be 100 times larger than that of $\kappa_{tqh}=0.01$, and the cross 
section of 50 TeV can be 9.1 (6.6) times larger than that of 7 TeV with a 60 (120) GeV electron beam. We also find that 
the cross section between polarized and unpolarized electron beam cases are related as:
$\sigma_{e^{-}_{r}}=\sigma_{e^{-}_{0}}\cdot(1-p_{e^{-}_{r}})$, $\sigma_{e^{-}_{l}}+\sigma_{e^{-}_{r}}=2\sigma_{e^{-}_{0}}$, 
independent of being case I,  II or III. Here $\sigma_{e^{-}_{r}}$, $\sigma_{e^{-}_{l}}$ and $\sigma_{e^{-}_{0}}$
represent the right, left and without electron beam polarization, respectively.

The cross section of the signal and backgrounds (in units of fb) are summarized in Table \ref{SBaftercuts_cut_7} (Cut-based method)
and Table \ref{SBaftercuts_tmva_7} (MVA-based method). From these tables, we calculate  
the signal significance $S/\sqrt{S+B}$ as 4.191 (15.341) and 6.652 (19.236) for 7 and 50 TeV by 
Cut-base method and 4.921 (16.934) and 7.874 (20.785) by MVA-based method after imposing all the relevant event selections (only the first
three selections in MVA-based method), respectively. Obviously, compared to the 
Cut-based method, the MVA-based method can get a better signal significance. As expected, with the 
$p_{2}$ = -0.6 e-beam polarization, the results are improved as 5.302 (19.404) and 8.414 (24.335) for Cut-based method
and 6.224 (21.420) and 9.960 (26.291) for MVA-based method. In addition to effective cuts, enhancing the b-tagging
efficiency together with reducing the jet mis-identification rates is one of the other way to improve the signal significance.
It is confirmed that the signal significance can be increased from 4.191, 6.652, 15.341 and 19.238 to 8.366, 13.840, 
33.750 and 44.154 with $\rm \epsilon_b=80$ $\%$, $\rm \epsilon_c=1$ $\%$, $\rm \epsilon_{light}=0.1$ $\%$ with the same
value of the input parameters and kinematic cuts.

\begin{table}[!htbp]

\caption{\label{SBaftercuts_cut_7}
Expected cross sections after all the selections for signal and backgrounds at the LHeC with an integrated luminosity of $\rm 1000\ fb^{-1}$,
b-tagging efficiency $\rm \epsilon_b$ = 60 $\%$, jet mis-identification rates $\rm \epsilon_c$ = 10 $\%$, $\rm \epsilon_{light}$ = 1 $\%$ by Cut-based method.
Especially, we select e-beam polarizations as $p_{0}$ = 0, $p_{1}$ = 0.6 and $p_{2}$ = -0.6.} 
\begin{center}
\begin{tabular}{c||  c  |  c   c   c}
\hline
\hline
\multirow{1}{*}{}  &  \multirow{1}{*}{}  &~~~ {S}~~~&~~~{B}~~~&~~~{SS}~~~    \\

\hline
                                                      & ~$p_{0}$~ &0.14 &0.93 &4.191  \\
60 GeV $\oplus$ 7 TeV $@$ LHeC    &$p_{1}$& 0.05 &0.37 & 2.651\\
                                                     &$p_{2}$ &0.22&1.49 &5.302  \\
\hline
                                                      &  $p_{0}$  &0.32&1.98 &6.652 \\
120 GeV $\oplus$ 7 TeV $@$ LHeC    &$p_{1}$& 0.13 &0.79 & 4.207\\
                                                     &$p_{2}$ &0.51 &3.16 &8.414  \\
\hline
                                                      &  $p_{0}$  &1.29 &5.80 &15.341 \\
60 GeV $\oplus$ 50 TeV $@$ LHeC    &$p_{1}$& 0.52&2.32 &9.702\\
                                                     &$p_{2}$ &2.07&9.28 &19.404  \\
\hline
                                                      &  $p_{0}$ &2.14 &10.26&19.238  \\
120 GeV $\oplus$ 50 TeV $@$ LHeC    &$p_{1}$&0.86&4.10& 12.167\\
                                                     &$p_{2}$ &3.43&16.42&24.335  \\
\hline
\hline
\end{tabular} 
\end{center}

\end{table}
\begin{table}[!htbp]

\caption{\label{SBaftercuts_tmva_7}
The same as Table \ref{SBaftercuts_cut_7} but for MVA-based method. We select e-beam polarizations as $p_{0}$ = 0, $p_{1}$ = 0.6 and $p_{2}$ = -0.6.}
\begin{center}
\begin{tabular}{c||  c  |  c   c   c}
\hline
\hline
\multirow{1}{*}{}  &  \multirow{1}{*}{}  &~~~ {S}~~~&~~~{B}~~~&~~~{SS}~~~    \\

\hline
                                                      & ~$p_{0}$ ~ &0.125 &0.520 &4.921  \\
60 GeV $\oplus$ 7 TeV $@$ LHeC    &$p_{1}$& 0.050 &0.208 & 3.112\\
                                                     &$p_{2}$ &0.200 &0.833 &6.224  \\
\hline
                                                      & $p_{0}$  &0.281&0.992 &7.874  \\
120 GeV $\oplus$ 7 TeV $@$ LHeC    &$p_{1}$& 0.112 &0.397 &4.980\\
                                                     &$p_{2}$ &0.450&1.588 &9.960  \\
\hline
                                                      &$p_{0}$  &0.652 &0.830 &16.934 \\
60 GeV $\oplus$ 50 TeV $@$ LHeC    &$p_{1}$& 0.261 &0.332 &10.710\\
                                                     &$p_{2}$ &1.043 &1.328 &21.420  \\
\hline
                                                      &$p_{0}$&1.082&1.629&20.785 \\
120 GeV $\oplus$ 50 TeV$@$ LHeC    &$p_{1}$&0.433 &0.652& 13.145\\
                                                     &$p_{2}$ &1.732&2.606&26.291  \\
\hline
\hline
\end{tabular} 
\end{center}

\end{table}

\begin{figure}[hbtp]
\centering
\includegraphics[scale=0.3]{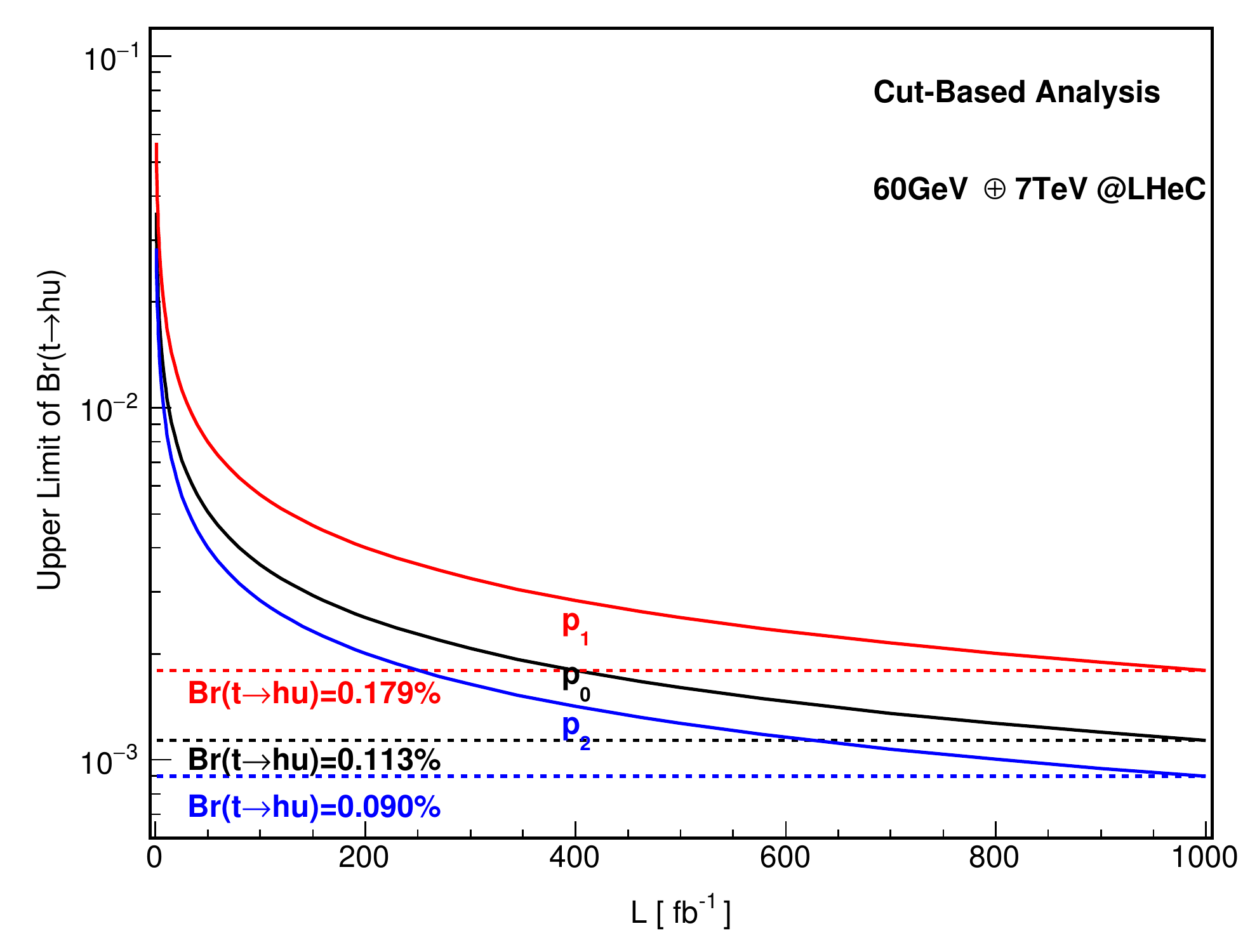}
\includegraphics[scale=0.3]{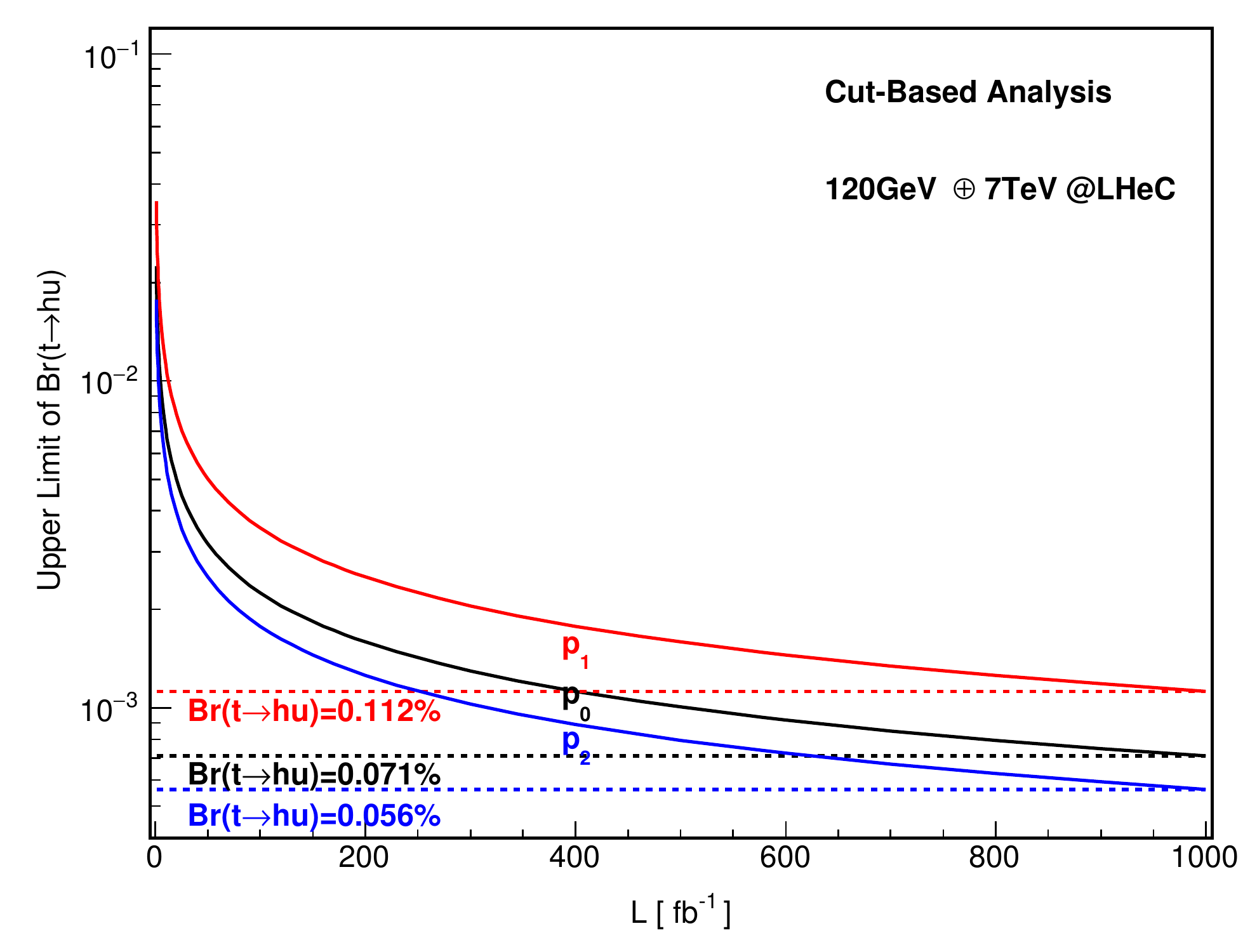}
\includegraphics[scale=0.3]{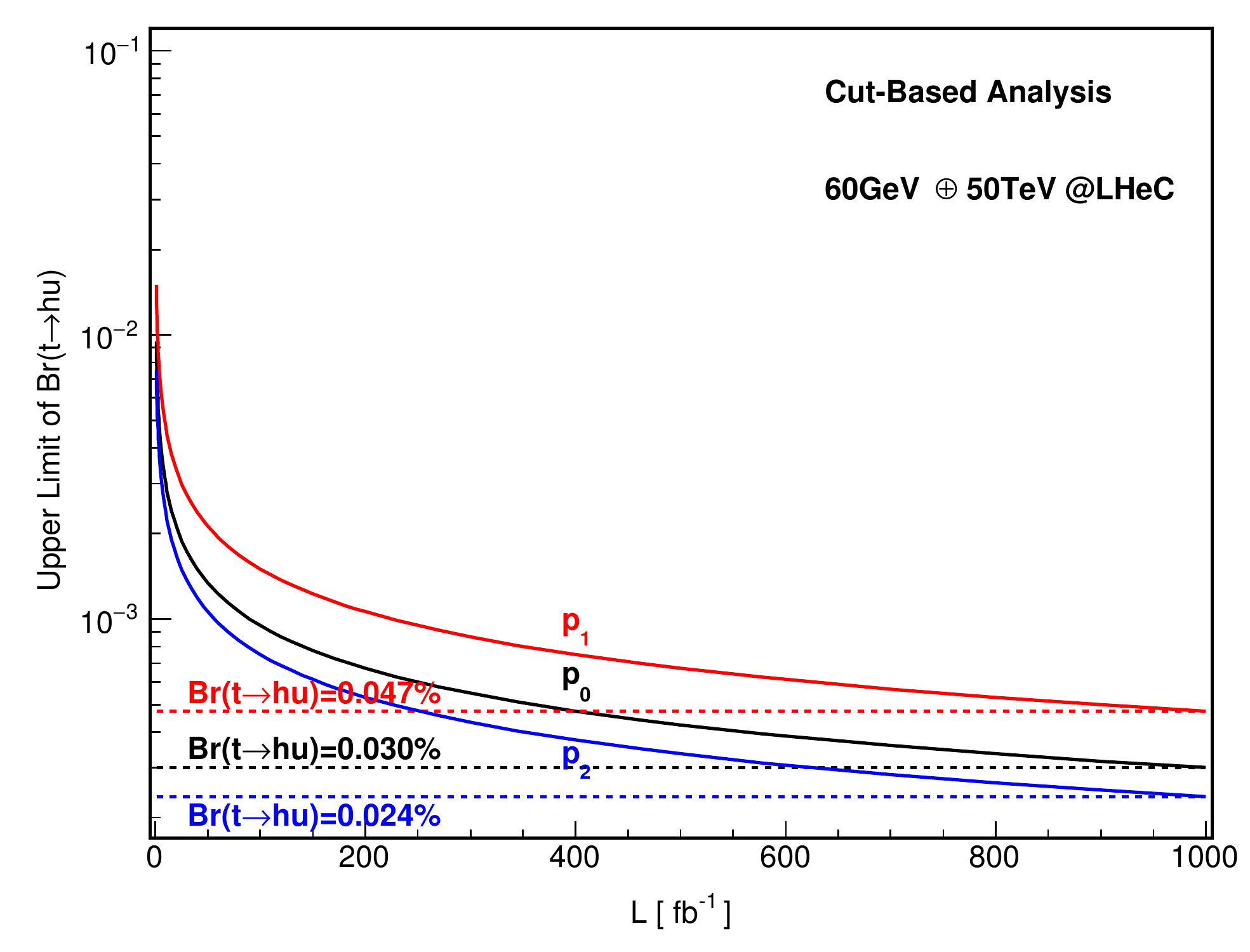}
\includegraphics[scale=0.3]{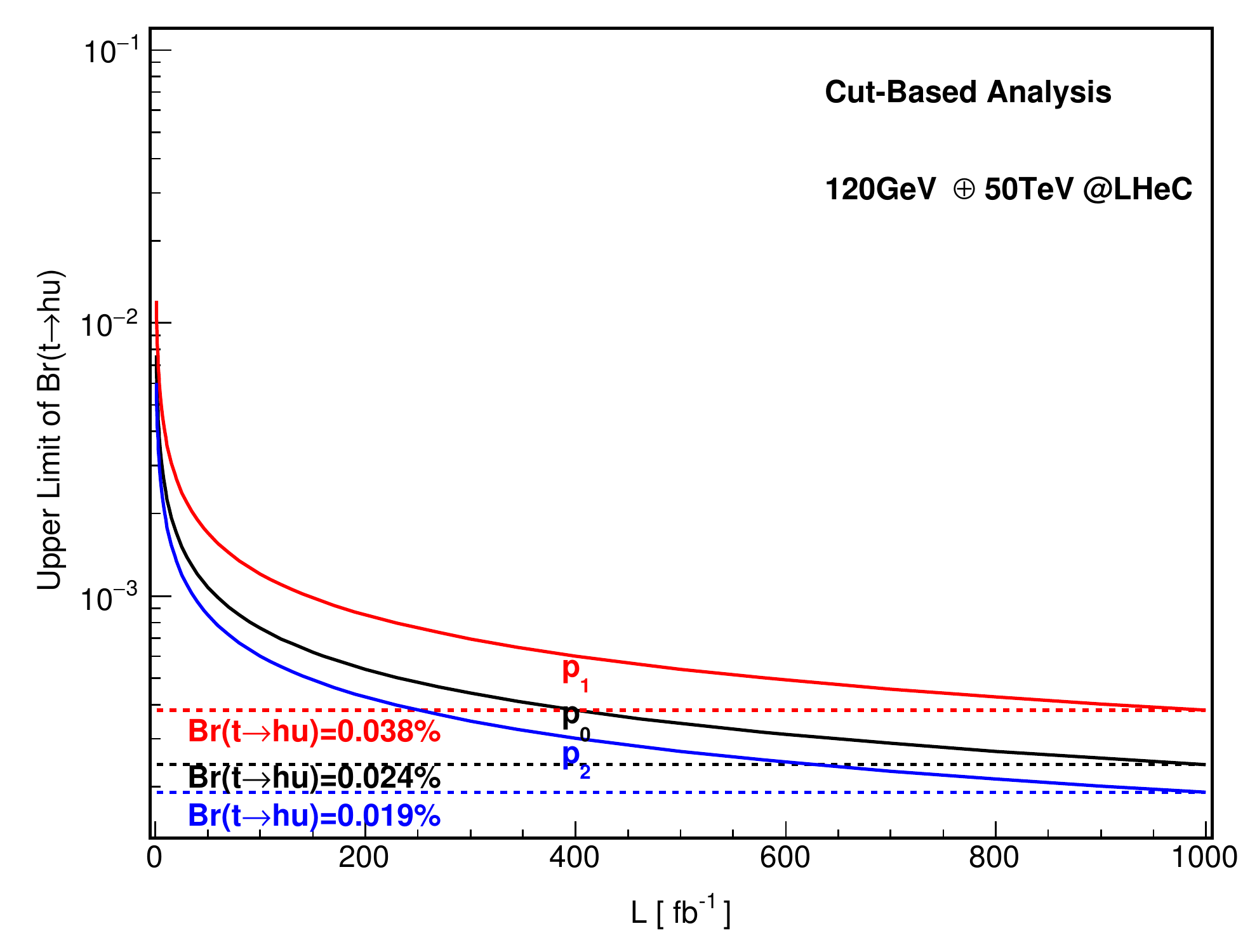}
\caption{\label{limit_cut_7}
The  upper limit from Cut-based method at 1$\sigma$ level at 7 (50) GeV LHeC with 60 (120) GeV electron beam. The red, blue and black curves
represent the 0.6, $-0.6$ and without electron beam polarization.}
\end{figure}

\begin{figure}[hbtp]
\centering
\includegraphics[scale=0.3]{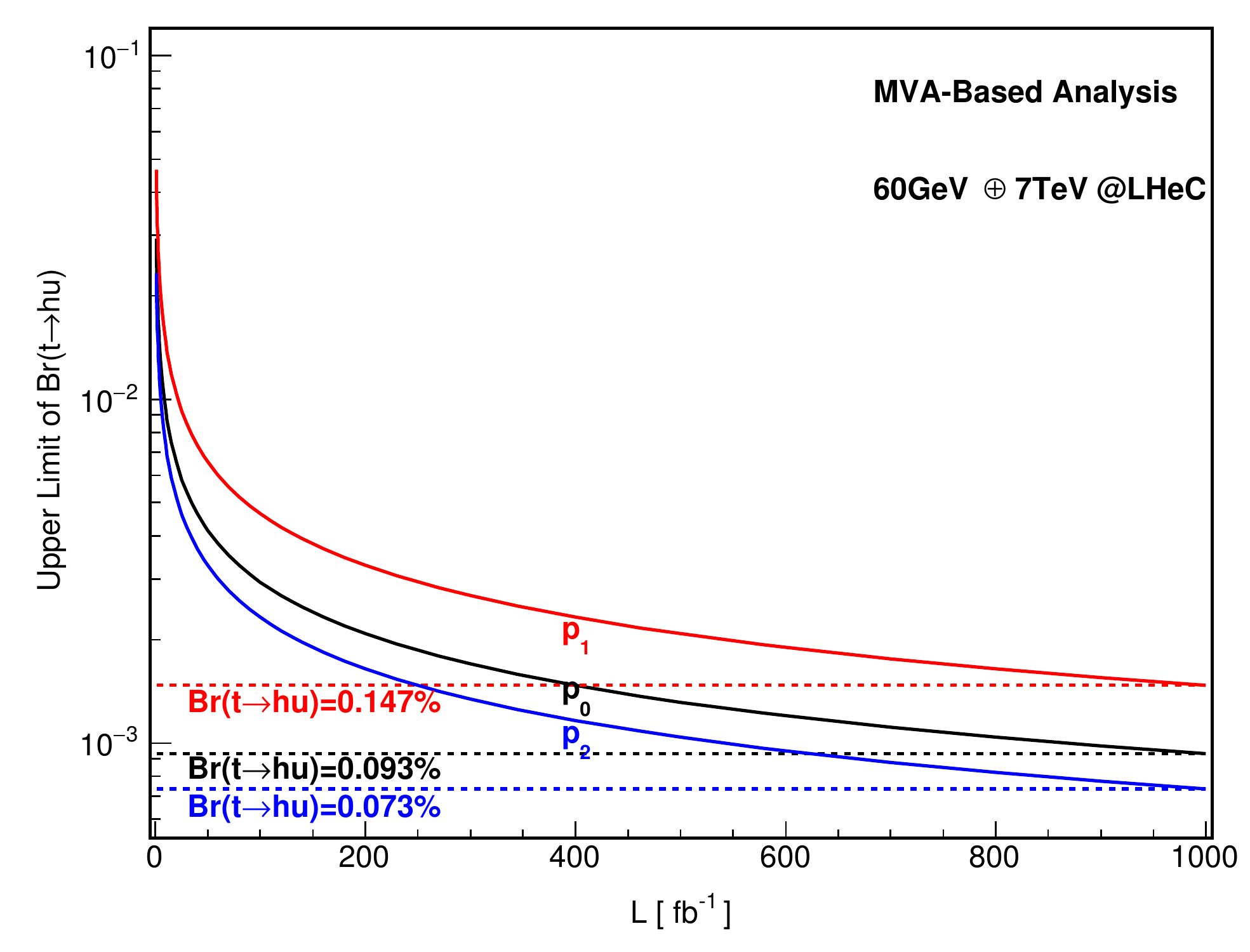}
\includegraphics[scale=0.3]{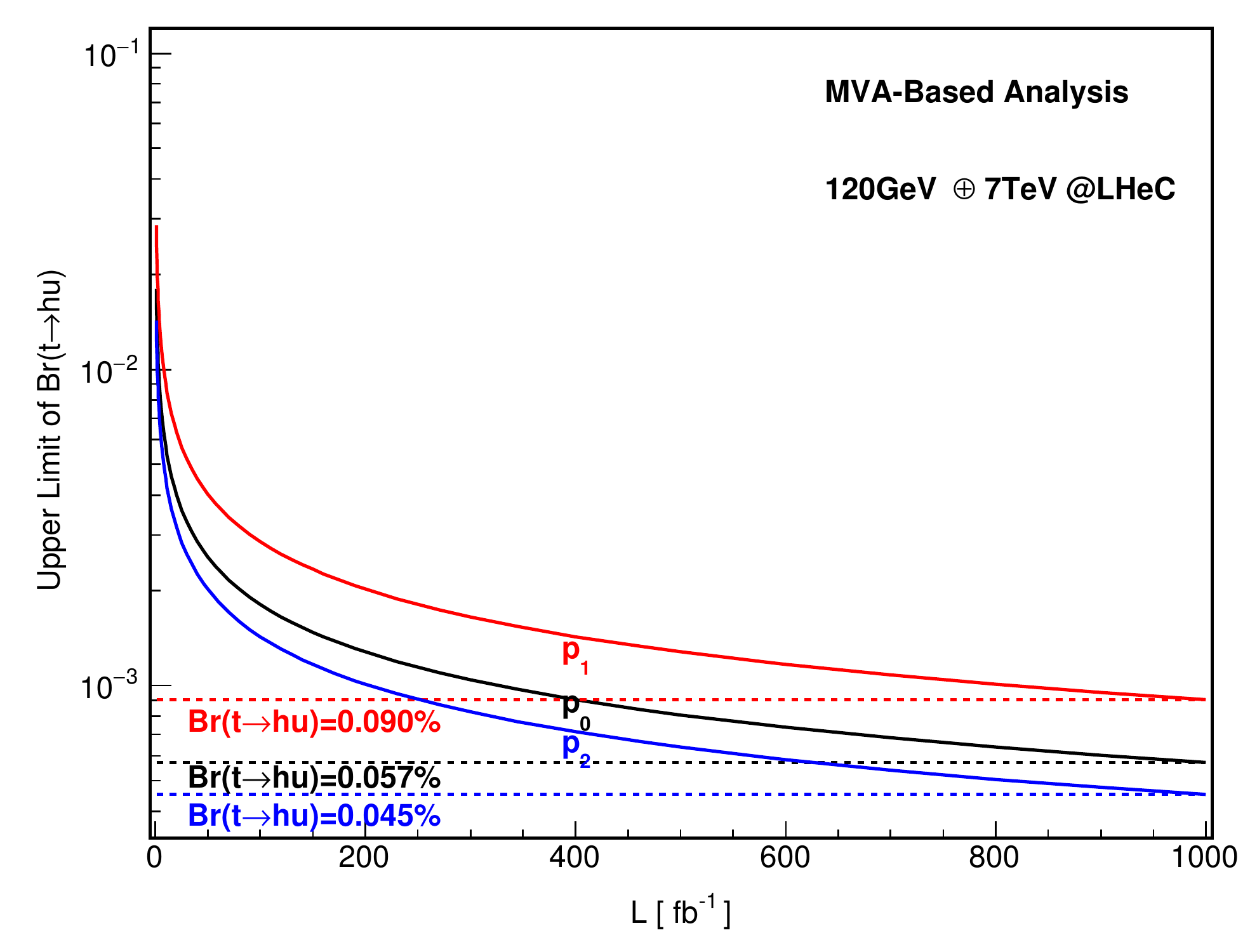}
\includegraphics[scale=0.3]{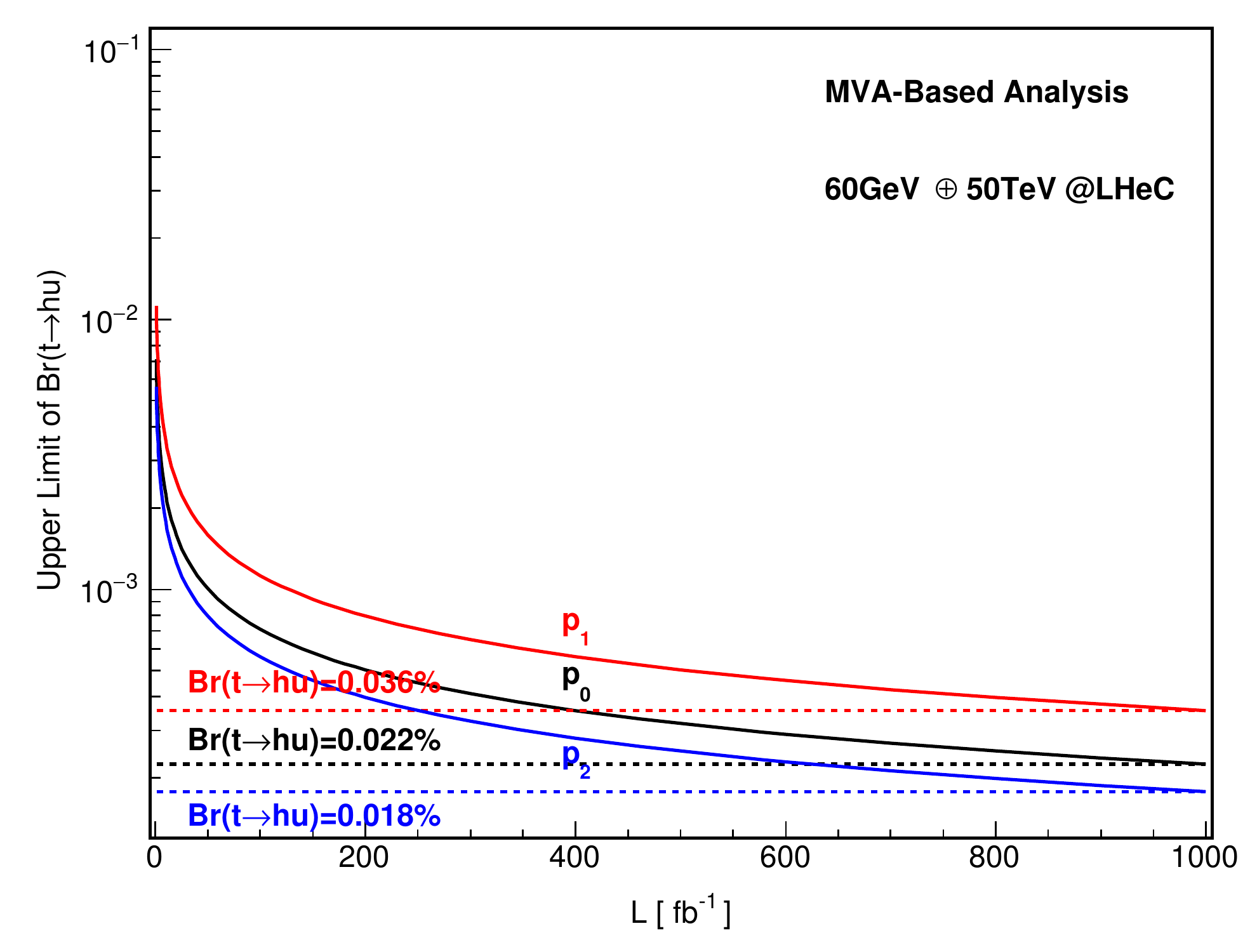}
\includegraphics[scale=0.3]{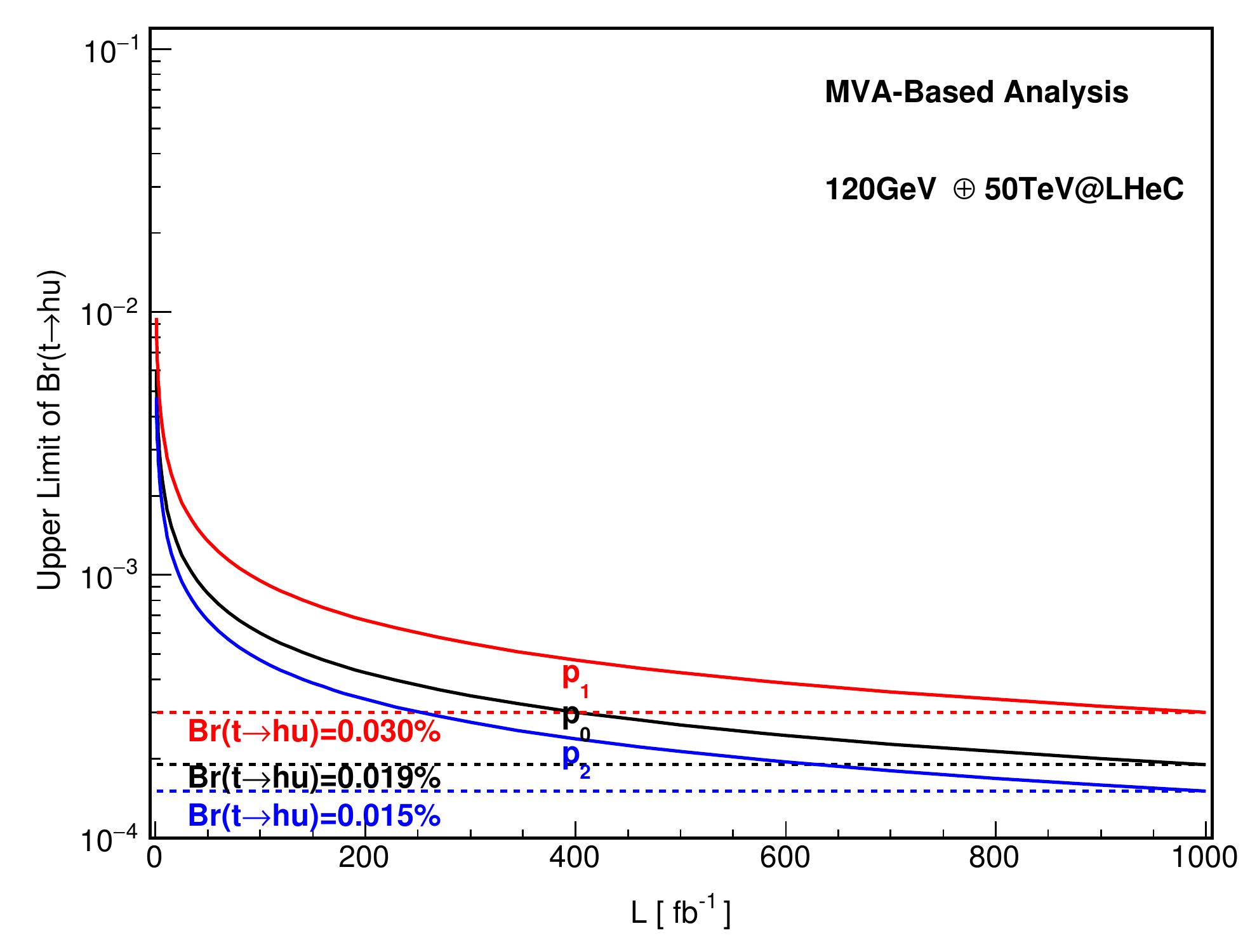}
\caption{\label{limit_tmva_7}
The same as Fig. \ref{limit_cut_7} but for MVA.}
\end{figure}

In order to estimate the sensitivity to the anomalous tqH couplings, we used chi-square ($\chi^2$) function \cite{Anomaloustqr, tqr}:
\begin{eqnarray}
\rm  \chi^2 = (\frac{\sigma_{tot}-\sigma_{B}}{\sigma_{B}\delta})^2
\end{eqnarray}
where $\rm\sigma_{tot}$ is the total cross section and $\rm\delta$ is the statistical error. In Fig. \ref{limit_cut_7} (Cut-based Analysis)
and Fig. \ref{limit_tmva_7} (MVA-based Analysis) \cite{TMVA}, we plot the contours of 1$\sigma$ limits to $\kappa_{tqH}$ at 7 (50) GeV LHeC
and 60 (120) GeV electron beam with different polarizations. The red, blue and black curves represent the 0.6, -0.6 and without 
electron beam polarization. From these figures, we can see that the branching ratio $\rm Br$ $(t\to uh)$ can be probed to 
$0.113$ (0.093) $\%$, $0.071$ (0.057) $\%$, $0.030$ (0.022) $\%$ and 0.024 (0.019) $\%$ with the Cut-based (MVA-based) Analysis at ($E_{p}$, $E_{e}$) = (7 TeV, 60 GeV),
($E_{p}$, $E_{e}$) = (7 TeV, 120 GeV),
($E_{p}$, $E_{e}$) = (50 TeV, 60 GeV) and ($E_{p}$, $E_{e}$) = (50 TeV, 120 GeV) beam energy.
As expected, the MVA-based method has a great advantage and also the 50 TeV high energy can get better results than 
the 7 TeV ones. Furthermore, it is clear that the limits can be probed down to 0.090 (0.073) $\%$, 0.056 (0.045) $\%$, 0.024 (0.018) $\%$ and 
0.019 (0.015) $\%$ with the e-beam polarization of $p_{2}$ = -0.6.

Finally, we give a precise integrated luminosity ($\cal L$) corresponding to the critical limits
obtained by the experimental results (Table \ref{luminosity_needed_Exp}) and other phenomenological studies (Table \ref{luminosity_needed_pheno})
. With the e-beam polarization $p_{2}$ = -0.6, the $\cal L$ needed to get the upper bounds on the $\rm Br$ $(t\to qh)$
 is reduced significantly. A detailed comparison between the LHeC collider(s) and the LHC or linear colliders are given.

\begin{table}[!htbp]
\caption{\label{luminosity_needed_Exp}
The integrated luminosity ($\cal L$) needed to get the upper bounds on the $\rm Br(t\to qh)$ at $95\%$ C.L. obtained from 
the experiments. Both the Cut (MVA) based results and 1$\sigma$ (2$\sigma$) limits with e-beam polarization are presented.}
\begin{center}
\vspace{0.2cm}
\begin{tabular}{c|| c | c c c | c c c}
\hline
\hline
\multirow{2}{*}{Channels and Limits}  &  \multirow{2}{*}{Method}  & \multicolumn{3}{c|}{$\rm {\cal L}[fb^{-1}]_{1\sigma}$}  & \multicolumn{3}{c}{ $\rm {\cal L}[fb^{-1}]_{2\sigma}$}  \\
\cline{3-8}
 & & $p_{0}$ & $p_{1}$  & $p_{2}$  &$p_{0}$ & $p_{1}$  & $p_{2}$  \\
\hline
$\rm t\bar{t}\to Wbqh \to \ell\nu b\gamma\gamma q$  &Cut &0.93 & 2.32 & 0.58 & 3.60&9.00&2.25\\
ATLAS, 4.7 (20.3) $\rm fb^{-1}$ @ 7 (8) TeV  &    & & &  & \\
$\rm Br$ $(t\to qh)< 0.79$ $\%$ \cite{FCNC_limit_tHq_ATLAS, FCNC_limit_tHq_ATLAS1} & MVA  &0.58 & 1.44 &0.36&2.24 & 5.60 & 1.40 \\
\hline
$\rm t\bar{t}\to Wbqh\to \ell\nu b\gamma\gamma q$&Cut &2.86  &7.15  &1.79 &11.10&27.76&6.94 \\
CMS, 19.5 $\rm fb^{-1}$ @ 8 TeV  &    & & &  & \\
$\rm Br$ $(t\to uh)< 0.45$ $\%$ \cite{FCNC_limit_tHq_CMS} & MVA&1.78  &4.45   & 1.11  &6.91&17.27&4.32  \\
\hline
$\rm D^0-\bar{D^0}$ mixing data                                &Cut &2.32&5.79&1.45&8.99&22.48&5.62  \\
$\rm Br$ $(t\to qh)<0.5$ $\%$ \cite{FCNC_limit_tHq_DDbar}  &MVA&1.44  &3.60 &0.90 &5.60&13.99&3.50  \\
\hline
$\rm Z\to c\bar{c}$ and EW observables                   &Cut   &13.13  & 32.83 &8.21  &51.01&127.53&31.88\\
$\rm Br$ $(t\to qh)<0.21$ $\%$ \cite{tqh_z2bb}       &MVA  &8.17 &20.43  & 5.11 & 31.74&79.35&19.84  \\
\hline
\hline
\end{tabular} 
\end{center}

\end{table}

\begin{table}[!htbp]
\caption{\label{luminosity_needed_pheno}
The same as Table \ref{luminosity_needed_Exp} but for some other phenomenological studies.}
\begin{center}
\vspace{0.2cm}
\begin{tabular}{c|| c | c c c | c c c}
\hline
\hline
\multirow{2}{*}{Channels and Limits}  &  \multirow{2}{*}{Method}  & \multicolumn{3}{c|}{$\rm {\cal L}[fb^{-1}]_{1\sigma}$}  & \multicolumn{3}{c}{ $\rm {\cal L}[fb^{-1}]_{2\sigma}$}  \\
\cline{3-8}
& & $p_{0}$ & $p_{1}$  & $p_{2}$  & $p_{0}$ & $p_{1}$  & $p_{2}$  \\
\hline
\hline
$\rm Wt\to Whq \to \ell\nu b\gamma\gamma q$          &Cut   &10.05&25.14&6.28 &39.05&97.63&24.41\\
LHC, 3000 $\rm fb^{-1}$ @ 14 TeV                         &      & & &  & \\
$3\sigma$, $\rm Br$ $(t\to qh)< 0.24$ $\%$ \cite{tqh_ppWhj}  & MVA &6.26 & 15.64 &3.91&24.30&60.75&15.19  \\
\hline
$\rm t\bar{t}\to Wbqh\to \ell\nu b\gamma\gamma q$ &Cut   &10.95&27.37&6.84 &42.52&106.31&26.58\\
LHC, 3000 $\rm fb^{-1}$ @ 14 TeV                     &      & & &  & \\
$\rm Br$ $(t\to uh)< 0.23$ $\%$ \cite{EFT_tqH_2}        & MVA &6.81 &17.03  & 4.26&26.46&66.15 & 16.54\\
\hline
$\rm t\bar{t}\to tqh\to \ell\nu bb\bar{b} q$ & Cut  &46.20&115.50&28.87&179.44&448.60&112.15\\
ILC, 3000 fb$^{-1}$ @ 500 GeV                &      & & &  & \\
$\rm Br$ $(t\to qh)< 0.112$ $\%$ \cite{ILC_tqh}  & MVA & 28.75 &71.86&17.97&111.65&279.13&69.78 \\
\hline
$\rm th\to \ell\nu b\tau^+\tau^-$           & Cut  & 25.75  &64.37  &16.09 &100.01&250.03&62.51 \\
LHC, 100 fb$^{-1}$ @ 13 TeV                    &      & & &  & \\
$\rm Br$ $(t\to uh)< 0.15$ $\%$ \cite{EFT_tqH_3}   & MVA & 16.02 & 40.05  &10.01 &62.23&155.58&38.89\\
\hline
$\rm th\to \ell\nu b\ell^+\ell^-X$          & Cut  &11.97 &29.92  &7.48  & 46.48&116.20&29.05 \\
LHC, 100 fb$^{-1}$ @ 13 TeV                    &      &  & &  & \\
 $\rm Br$ $(t\to uh)< 0.22$ $\%$ \cite{EFT_tqH_3}  & MVA & 7.45 &18.61  &4.65&28.92&72.30&18.08 \\
\hline
$\rm th\to jjb b\bar{b}$                    & Cut  &4.47 &11.17  &2.79& 17.35&43.38&10.84 \\
LHC, 100 fb$^{-1}$@13TeV                    &      & & &  & \\
$\rm Br$ $(t\to uh)< 0.36$ $\%$ \cite{EFT_tqH_3}   & MVA &2.78 &6.95   &1.74& 10.80&26.99&6.75  \\

\hline
\hline
\end{tabular} 
\end{center}

\end{table}
\section{Conclusion}
In this paper, we investigated the anomalous FCNC Yukawa interactions between the top quark, the Higgs boson, and either an up or 
charm quark with a channel $\rm e^- p\rightarrow \nu_e \bar{t} \rightarrow \nu_e h \bar{q}(h\rightarrow b\bar{b})$  at the LHeC.
The signal significance $S/\sqrt{S+B}$ can be obtained as 4.191 (4.921), 6.652 (7.874), 15.341 (16.934) and 19.238 (20.785) with the Cut-based (MVA-based)
method at ($E_{p}$, $E_{e}$) = (7 TeV, 60 GeV), ($E_{p}$, $E_{e}$) = (7 TeV, 120 GeV), ($E_{p}$, $E_{e}$) = (50 TeV, 60 GeV) and ($E_{p}$, $E_{e}$) = (50 TeV, 120 GeV). Similarly, our results show that the branching ratio $\rm Br$ $(t\to uh)$ can be probed to 
 0.113 (0.093) $\%$, 0.071 (0.057) $\%$, 0.030 (0.022) $\%$ and 0.024 (0.019) $\%$, and with the e-beam polarization $p_{2}=-0.6$, 
 the expected limits can be greatly reduced. Finally, a detailed comparison between our study and the critical limits obtained by the experiments
 and other phenomenological studies are shown. We thus give an overview of the search potential on the anomalous top-Higgs couplings
 with polarized electron beam at the LHeC.
\section*{Acknowledgments} \hspace{5mm}
Project supported by the National Natural Science Foundation of China
(Grant No. 11675033), by the Fundamental Research Funds for the
Central Universities (Grant No. DUT15LK22).


\begin{thebibliography}{99}
\expandafter\ifx\csname
natexlab\endcsname\relax\def\natexlab#1{#1}\fi
\expandafter\ifx\csname bibnamefont\endcsname\relax
  \def\bibnamefont#1{#1}\fi
\expandafter\ifx\csname bibfnamefont\endcsname\relax
  \def\bibfnamefont#1{#1}\fi
\expandafter\ifx\csname citenamefont\endcsname\relax
  \def\citenamefont#1{#1}\fi
\expandafter\ifx\csname url\endcsname\relax
  \def\url#1{\texttt{#1}}\fi
\expandafter\ifx\csname
urlprefix\endcsname\relax\def\urlprefix{URL }\fi
\providecommand{\bibinfo}[2]{#2}
\providecommand{\eprint}[2][]{\url{#2}}

\bibitem{LHeC_design_Note}
Oliver Br$\ddot{u}$ening, Max Klein,
{\it The Large Hadron Electron Collider},
Mod.Phys.Lett.A, Vol.28, No.16 (2013) 1330011,
LHeC-Note-2013-001 GEN, [arXiv:1305.2090].

\bibitem{FCC-he_Study}
M.Klein, {\it Development of the FCC-he Study},
In FCC Physics, Detector and Accelerator Workshop,
Istanbul, March 2016.

\bibitem{twb_NLO}
C. S. Li, R. J. Oakes, and T. C. Yuan,
{\it QCD corrections to $t\rightarrow W^+b$ }, Phys. Rev. D 43 (1991) 3759-3762.

\bibitem{decay_tqH}
Wei-Shu Hou, {\it Tree level $t \rightarrow ch$ or $h \rightarrow  t {\bar c}$ decays},
Phys. Lett. B 296 (1992) 179-184.

\bibitem{FCNC_limit_tHq_ATLAS}
G. Aad et al., [ATLAS Collaboration], {\it Search for top quark decays $t\rightarrow qH$
with $H\rightarrow \gamma\gamma$ using the ATLAS detector},
JHEP 06 (2014) 008, CERN-PH-EP-2014-036 (2014), [arXiv:1403.6293].

\bibitem{FCNC_limit_tHq_ATLAS1}
{\it Search for flavour changing neutral currents in top quark decays
$t\rightarrow cH$, with $H\rightarrow \gamma\gamma$,
and limit on the tcH coupling with the ATLAS detector at the LHC},
ATLAS-CONF-2013-081 (2013).

\bibitem{FCNC_limit_tHq_CMS}
[CMS Collaboration], {\it Searches for heavy Higgs bosons
in two-Higgs-doublet models and
for $t\rightarrow ch$ decay using multilepton
and diphoton final states in pp collisions at 8 TeV},
Phys. Rev. D 90 (2014) 112013, CMS-HIG-13-025, CERN-PH-EP-2014-239, [arXiv:1410.2751].
[CMS Collaboration], {\it Combined multilepton and diphoton limit on $t\rightarrow cH$},
CMS-PAS-HIG-13-034 (2014).

\bibitem{FCNC_limit_tHq_DDbar}
J. I.  Aranda, A. Cordero-Cid, F. Ramirez-Zavaleta, J.J. Toscano, and E.S. Tututi,  	
{\it Higgs mediated flavor violating top quark decays $t\to u_{i}H, u_i\gamma, u_i\gamma\gamma$,
and the process $\gamma\gamma\to tc$ in effective theories},
Phys. Rev. D 81 (2010) 077701, [arXiv:0911.2304].

\bibitem{tqh_z2bb}
F. Larios, R. Martinez, M.A. Perez
{\it Constraints on top quark FCNC from electroweak precision measurements},
Phys.Rev. D72 (2005) 057504, [arXiv:hep-ph/0412222].

\bibitem{tqh_ppWhj}
Yao-Bei Liu, Zhen-Jun Xiao,
{\it Searches for top-Higgs FCNC couplings via Whj signal with $h\to\gamma\gamma$ at the LHC}, [arXiv:1605.01179].

\bibitem{EFT_tqH_2}
Lei Wu,  {\it Enhancing thj Production from Top-Higgs FCNC Couplings},
JHEP 02 (2015) 061, [arXiv:1407.6113].

\bibitem{ILC_tqh}
Hoda Hesari, Hamzeh Khanpour, Mojtaba Mohammadi Najafabadi,
{\it Direct and Indirect Searches for Top-Higgs FCNC Couplings},
Phys.Rev. D92 (2015) 11, 113012, [arXiv:1508.07579].

\bibitem{EFT_tqH_3}
Admir Greljo, Jernej F. Kamenik, and Joachim Kopp,
{\it Disentangling Flavor Violation in the Top-Higgs Sector at the LHC},
JHEP 1407 (2014) 046, [arXiv:1404.1278].

\bibitem{FeynRules2.0}
A. Alloul, N. D. Christensen, C. Degrande, C. Duhr, and B.Fuks,
{\it FeynRules 2.0 - A complete toolbox for tree-level phenomenology},
Comput. Phys. Commun. 185, 2250-2300 (2014), [arXiv:1310.1921].

\bibitem{MadGraph5}
J. Alwall, R. Frederix, S. Frixione, V. Hirschi, F. Maltoni, O. Mattelaer, H.-S. Shao, T. Stelzer, P. Torrielli, and M. Zaro,
{\it The automated computation of tree-level and next-to-leading order differential cross sections, and their
matching to parton shower simulations}, JHEP 1407, 079 (2014), [arXiv:1405.0301].

\bibitem{Pythia6.4}
T. Sjostrand, S. Mrenna, and P. Z. Skands, {\it PYTHIA 6.4 Physics and Manual},
JHEP 0605, 026 (2006), [hep-ph/0603175].

\bibitem{CTEQ6L}
J. Pumplin, D. R. Stump, J. Huston, H.L. Lai, P. M. Nadolsky, and W.K. Tung,
{\it New generation of parton distributions with uncertainties from global QCD analysis},
JHEP 0207 (2002) 012, [arXiv:hep-ph/0201195];
D. Stump, J. Huston, J. Pumplin, W.-K. Tung, H.L. Lai, S. Kuhlmann, and J.F. Owens,
{\it Inclusive jet production, parton distributions, and the search for new physics},
JHEP 0310 (2003) 046.

\bibitem{TMVA}
A. Hoecker, et.al., {\it TMVA - Toolkit for Multivariate Data Analysis},
PoS ACAT (2007) 040, CERN-OPEN-2007-007, [arXiv:physics/0703039].

\bibitem{Anomaloustqr}
M.K\"{o}ksal, S. C. Inan, {\it Anomalous $tq\gamma$ couplings in
$\gamma p$ collision at the LHC},
Advances in High Energy Physics, 935840(2014), [arXiv:1305.7096].

\bibitem{tqr}
Hao Sun, {\it Probe Anomalous $tq\gamma$ couplings through Single Top
Photoproduction at the LHC}, Nucl. Phys. B 886 (2014) 691-711, [arXiv:1402.1817].
\end{thebibliography}
\end{document}